\documentclass[a4paper,11pt]{article}
\pdfoutput=1 

\usepackage{jcappub} 

\usepackage{appendix}

\title{Revisit of the interacting holographic dark energy model after Planck 2015}

\author[a]{Lu Feng,}
\author[a,b,1]{Xin Zhang\note{Corresponding author.}}
\affiliation[a]{Department of Physics, College of Sciences, Northeastern University, \\Shenyang
110004, China}
\affiliation[b]{Center for High Energy Physics, Peking University, \\Beijing 100080, China}

\emailAdd{fengluu@foxmail.com}
\emailAdd{zhangxin@mail.neu.edu.cn}

\abstract{We investigate the observational constraints on the interacting holographic dark energy model. We consider five typical interacting models with the interaction terms $Q=3\beta H\rho_{\rm{de}}$, $Q=3\beta H\rho_{\rm{c}}$, $Q=3\beta H(\rho_{\rm{de}}+\rho_{\rm c})$, $Q=3\beta H\sqrt{\rho_{\rm{de}}\rho_{\rm c}}$, and $Q=3\beta H\frac{\rho_{\rm{de}}\rho_{c}}{\rho_{\rm{de}}+\rho_{\rm c}}$, respectively, where $\beta$ is a dimensionless coupling constant. The observational data we use in this paper include the JLA compilation of type Ia supernovae data, the Planck 2015 distance priors data of cosmic microwave background observation, the baryon acoustic oscillations measurements, and the Hubble constant direct measurement. We make a comparison for these five interacting holographic dark energy models by employing the information criteria, and we find that, within the framework of holographic dark energy, the $Q=3\beta H\frac{\rho_{\rm{de}}\rho_{\rm c}}{\rho_{\rm{ de}}+\rho_{\rm c}}$ model is most favored by current data, and the $Q=3\beta H\rho_{\rm c}$ model is relatively not favored by current data. For the $Q=3\beta H\rho_{\rm{de}}$ and $Q=3\beta H\frac{\rho_{\rm{de}}\rho_{\rm c}}{\rho_{\rm{ de}}+\rho_{\rm c}}$ models, a positive coupling $\beta$ can be detected at more than 2$\sigma$ significance.}

\begin{document}
\maketitle
\flushbottom

\section{Introduction}
\label{sec1}
The cosmological observations of type Ia supernovae (SNIa) \cite{Riess:1998cb,Perlmutter:1998np}, the cosmic microwave background (CMB) \cite{Spergel:2003cb,Bennett:2003bz}, and the large scale structure (LSS) \cite{Tegmark:2003ud,Abazajian:2004aja} have confirmed that our universe is undergoing an accelerating expansion. This strongly indicates the existence of dark energy \cite{Peebles:2002gy,Bean:2005ru,Copeland:2006wr,Sahni:2006pa,Kamionkowski:2007wv,Li:2011sd,Bamba:2012cp}, a mysterious exotic energy component with negative pressure, and the dark energy contributes about 70\% of the cosmic energy density. Hitherto, although lots of efforts have been made to understanding dark energy, its physical nature is still a mystery .

The most important theoretical candidate of dark energy is the cosmological constant $\Lambda$, which fits the observational data quite well. However, the cosmological constant model ($\Lambda$CDM) is plagued with the ``fine-tuning" and the ``cosmic coincidence" problems \cite{Weinberg:1988cp,Sahni:1999gb,Frieman:2008sn}, which implies that novel idea needs to be introduced to solve the theoretical problems in the $\Lambda$CDM model. The holographic dark energy (HDE) model \cite{Li:2004rb} was thus put forward, which is based on the holographic principle of quantum gravity theory and the effective quantum field theory. In this model, the vacuum energy is viewed as dark energy, and the holographic principle leads to the ultraviolet (UV) cutoff linked to the infrared (IR) cutoff of the effective quantum field theory in a subtle way. In order not to make the effective field theory breakdown in the presence of gravity, the theory requires that the total energy of a system with size $L$ should not exceed the mass of a black hole with the same size, i.e., $L^3\rho_{\rm de}\leq LM^2_{\rm pl}$ \cite{Cohen:1998zx}. In this way, we have the holographic dark energy density,
\begin{equation}\label{1.1}
 \rho_{\rm de}=3c^2 M^2_{\rm pl} L^{-2},
\end{equation}
where $c$ is a dimensionless model parameter, $M_{\rm pl}=\frac{1}{\sqrt{8\pi G}}$ is the reduced Planck mass, and $L$ is the IR cutoff size in the theory. Li \cite{Li:2004rb} suggested that the IR length-scale cutoff $L$ should be chosen to be the future event horizon of the universe, defined as
\begin{equation}\label{1.2}
L=a(t)\int_{t}^{\infty}\frac{dt'}{a(t')}=a\int_{a}^{\infty}\frac{da'}{Ha'^2},
\end{equation}
where $a(t)$ is the scale factor of our universe and $H$ is the Hubble parameter, $H=\dot{a}/a$, where the dot denotes the derivative with respect to $t$. Li's choice not only gives a reasonable value for the energy density of dark energy, but also leads to an accelerated universe. Moreover, the cosmic coincidence problem can also be explained successfully in this model once the inflation is also considered (see \cite{Li:2004rb} for details).

During the last decade, the holographic dark energy model has been studied widely \cite{Huang:2004wt,Wang:2004nqa,Huang:2004mx,Zhang:2005hs,Chang:2005ph,Nojiri:2005pu,Zhang:2006av,Zhang:2006qu,Zhang:2007sh,Ma:2007pd,Zhang:2007es,Zhang:2007an,Li:2008zq,Ma:2007av,Li:2009bn,Zhang:2009xj,Wang:2012uf,Li:2013dha,Zhang:2015rha,Cui:2015oda,delCampo:2011jp,Landim:2015hqa}. By far, various observational constraints on this model all indicate that the parameter $c$ is less than 1, implying that the holographic dark energy would lead to a phantom universe with big rip as its ultimate fate \cite{Zhang:2005yz}. One way of avoiding the big rip is to consider some phenomenological interaction between holographic dark energy and dark matter \cite{Li:2009zs,Zhang:2012uu}. With the help of the interaction, the big rip might be avoided due to the occurence of an attractor solution in which the effective equations of state of dark energy and dark matter become identical in the future.

In this paper, we will explore the possible phenomenological interaction between holographic dark energy and dark matter by using the latest observational data. We wish to see whether some hint of the existence of the direct coupling between dark energy and dark matter can be found in the HDE model after the 2015 data release of the Planck mission.

This paper is organized as follows. In section \ref{sec2}, we describe the interacting holographic dark energy model in a flat universe. In section \ref{sec3}, we introduce the analysis method and the observational data. The results are given and discussed in section \ref{sec4}. A summary is given in section \ref{sec5}.

\section{The interacting holographic dark energy model}\label{sec2}

In this section, we derive the basic equations for the interacting holographic dark energy (IHDE) model in a flat universe.

In a spatially flat Friedmann-Roberston-Walker universe, the Friedmann equation can be written as
\begin{equation}
3M^2_{\rm{pl}} H^2=\rho_{\rm c}+\rho_{\rm b}+\rho_{\rm r}+\rho_{\rm{de}},
\end{equation}
where $\rho_{\rm c}$, $\rho_{\rm b}$, $\rho_{\rm r}$, and $\rho_{\rm{de}}$ represent the energy densities of cold dark matter, baryon, radiation, and dark energy, respectively. For convenience, we define the fractional energy densities of various components, $\Omega_{i}=\rho_{i}/(3M^2_{\rm{pl}} H^2)$, where $3M^2_{\rm{pl}} H^2$ is the critical density of the universe. By definition, we have
\begin{equation}\label{2.2}
\Omega_{\rm{c}}+\Omega_{\rm{b}}+\Omega_{\rm r}+\Omega_{\rm{de}}=1.
\end{equation}
When we consider the direct, non-gravitational interaction between the two dark components, the conservation equations for all components can be written as
\begin{equation}\label{2.3}
\dot{\rho}_{\rm c}+3H\rho_{\rm c}=Q,
\end{equation}
\begin{equation}\label{2.4}
\dot{\rho}_{\rm de}+3H(\rho_{\rm de}+p_{\rm de})=-Q,
\end{equation}
\begin{equation}\label{2.5}
\dot{\rho_{\rm b}}+3H\rho_{\rm b}=0,
\end{equation}
\begin{equation}\label{2.6}
\dot{\rho_{\rm r}}+4H\rho_{\rm r}=0,
\end{equation}
where $Q$ denotes the phenomenological interaction term. In this work, we consider five cases for the interaction term $Q$,
\begin{equation}
Q_1=3\beta H\rho_{\rm{de}},
\end{equation}
\begin{equation}
Q_2=3\beta H\rho_{\rm c},
\end{equation}
\begin{equation}
Q_3=3\beta H(\rho_{\rm{de}}+\rho_{\rm c}),
\end{equation}
\begin{equation}
Q_4=3\beta H\sqrt{\rho_{\rm de}\rho_{\rm c}},
\end{equation}
\begin{equation}
Q_5=3\beta H\frac{\rho_{\rm de}\rho_{\rm c}}{\rho_{\rm de}+\rho_{\rm c}}.
\end{equation}
These interaction forms have all been widely studied; see, e.g., \cite{Li:2009zs,Zhang:2012uu,Barrow:2006hia,Zhang:2005rg,Zhang:2005rj,Valiviita:2008iv,Zhang:2007uh,Zhang:2009qa,Li:2010ak,Clemson:2011an,Zhang:2013lea,Bolotin:2013jpa,Costa:2013sva,Li:2013bya,yang:2014vza,Li:2014eha,Li:2014cee,Funo:2014poa,Geng:2015ara,Valiviita:2015dfa,Bouhmadi-Lopez:2016dcs,Li:2015vla,Wang:2016lxa,Nunes:2016dlj,Sola:2016ecz}.
Note that, according to our convention, $\beta>0$ means that dark energy decays to dark matter, and $\beta<0$ means that dark matter decays into dark energy. Usually, $\beta<0$ will lead to unphysical consequences in physics, i.e., $\rho_{\rm c}$ will become negative and $\Omega_{\rm{de}}$ will become greater than 1 in the far future.

Combining eqs. (\ref{2.2})--(\ref{2.6}) gives
\begin{equation}
p_{\rm de}=-\frac{2}{3}\frac{\dot{H}}{H^2}\rho_{\rm c}-\rho_{\rm c}-\frac{1}{3}\rho_{\rm r},
\end{equation}
which together with energy conservation equation (\ref{2.4}) for dark energy leads to
\begin{equation}\label{2.13}
2\frac{\dot{H}}{H}(\Omega_{\rm{de}}-1)+\dot{\Omega}_{\rm{de}}+H(3\Omega_{\rm{de}}+\Omega_{\rm I}-3-\Omega_{\rm r})=0,
\end{equation}
where
\begin{equation}
 \Omega_{\rm I}=\frac{Q}{3M_{\rm{pl}}^2H^3}.
\end{equation}
From the holographic dark energy density equation (\ref{1.1}), we have
\begin{equation}
L=\frac{c}{H\sqrt{\Omega_{\rm{de}}}},
\end{equation}
and then we have
\begin{equation}\label{2.16}
r(t)=\frac{L}{a}=\frac{c}{Ha\sqrt{\Omega_{\rm{de}}}}.
\end{equation}
Combining eq. (\ref{1.2}) with eq. (\ref{2.16}) and taking derivative with respect to $t$, one can get
\begin{equation}\label{2.17}
\frac{\dot{\Omega}_{\rm{de}}}{2\Omega_{\rm{de}}}+H+\frac{\dot{H}}{H}=\frac{H}{c}\sqrt{\Omega_{\rm{de}}}.
\end{equation}
Combining eqs. (\ref{2.13}) and (\ref{2.17}), we finally have the following two equations governing the dynamical evolution of the interacting holographic dark energy in a flat universe,
\begin{equation}
\frac{1}{E}\frac{dE}{dz}=-\frac{\Omega_{\rm{de}}}{1+z}\left(\frac{1}{c}\sqrt{\Omega_{\rm{de}}}+\frac{1}{2}+\frac{\Omega_{\rm I}-3-\Omega_{\rm r}}{2\Omega_{\rm{de}}}\right),
\end{equation}
\begin{equation}
\frac{d\Omega_{\rm{de}}}{dz}=-\frac{2(1-\Omega_{\rm{de}})\Omega_{\rm{de}}}{1+z}\left(\frac{1}{c}\sqrt{\Omega_{\rm{ de}}}+\frac{1}{2}+\frac{\Omega_{\rm I}-\Omega_{\rm r}}{2(1-\Omega_{\rm{de}})}\right),
\end{equation}
where $E(z)=H(z)/H_0$ is the dimensionless Hubble expansion rate, and the fractional density of radiation $\Omega_{\rm{r}}(z)=\Omega_{\rm{r0}}(1+z)^4/E(z)^2$. In addition, we have $\Omega_{ \rm{r0}}=\Omega_{\rm{m0}} / (1+z_{\rm eq})$ with $z_{\rm eq}=2.5\times 10^4 \Omega_{\rm{m0}} h^2 (T_{\rm cmb}/2.7\,{\rm K})^{-4}$, where $\Omega_{\rm{r0}}$ and $\Omega_{\rm{m0}}$ are the present-day fractional energy densities of radiation and matter, respectively. Here, we take $T_{\rm cmb}=2.7255\,{\rm K}$, and $h$ is the dimensionless Hubble constant defined by $H_0=100h$ km s$^{-1}$ Mpc$^{-1}$.
The initial conditions of these two differential equations are $E_0=1$ and $\Omega_{\rm{de0}}=1-\Omega_{\rm{m0}}-\Omega_{\rm{r0}}$ at $z=0$. When we solve the equations, $\Omega_{\rm{m0}}$ and $c$ are free model parameters.

\section{Data and method}\label{sec3}

We study the cosmological constraints on the IHDE models with the most recent observational data. There are four free parameters, $c$, $h$, $\Omega_{m0}$, and $\beta$, in the IHDE models. For comparison, the fitting results of the original HDE model will also be presented.

We apply the $\chi^2$ statistic to estimate the model parameters. For each data set, we calculate $\chi^2_\xi=(\xi^{\rm obs}-\xi^{\rm th})^2/\sigma^2_\xi$, where $\xi$ is a physical quantity, $\xi^{\rm obs}$ is experimentally measured value, $\xi^{\rm th}$ is the theoretically predicted value, and $\sigma_{\xi}$ is the standard deviation.

The total $\chi^2$ is the sum of all $\chi^2_\xi$, i.e.,
\begin{equation}
\chi^2=\sum\limits_{\xi} \chi^2_\xi.
\end{equation}
In this paper, we perform a joint SN+CMB+BAO+$H_0$ fit, where the total $\chi^2$ is given by
\begin{equation}
  \chi^2=\chi^2_{\rm SN}+\chi^2_{\rm CMB}+\chi^2_{\rm BAO}+\chi^2_{H_0}.
\end{equation}

For comparing different models, a proper analysis method must be chosen. The $\chi^2$ comparsion is the simplest one which is widely used. However, for models with different number of parameters, the comparison using $\chi^2$ may be unfair. Therefore, we choose to use two information criteria: the Akaike information criterion (AIC) \cite{AIC1974} and the Bayesian information criterion (BIC) \cite{BIC1978}. They are defined as ${\rm BIC}=\chi^2_{\rm{min}}+k\ln N$ and ${\rm AIC}=\chi^2_{\rm{min}}+2k$, where $k$ is the number of parameters, and $N$ is the number of data points used in the fit. Actually, the relative value between different models for the information criteria needs to be paid more interest. Thus, we use $\Delta {\rm AIC}=\Delta\chi^2_{\rm{min}}+2\Delta k$ and $\Delta{\rm BIC}=\Delta\chi^2_{\rm{min}}+\Delta k\ln N$ for comparing models. In this paper, we choose the $\Lambda$CDM as a reference model.

We investigate five IHDE models in this paper. For convenience, in the following, the model with $Q_1=3\beta H\rho_{\rm{de}}$ is denoted as IHDE1, the model with $Q_2=3\beta H\rho_{\rm c}$ is denoted as IHDE2, the model with
$Q_3=3\beta H(\rho_{\rm{de}}+\rho_{\rm c})$ is denoted as IHDE3, the model with $Q_4=3\beta H\sqrt{\rho_{\rm{ de}}\rho_{\rm c}}$ is denoted as IHDE4, and the model with $Q_5=3\beta H\frac{\rho_{\rm{de}}\rho_{\rm c}}{\rho_{\rm{ de}}+\rho_{\rm c}}$ is denoted as IHDE5.

\subsection{Type Ia supernovae}

For the SN Ia data, we use the Joint-Light-curve Analysis (JLA) data compilation consisting of 740 SN data points \cite{Betoule:2014frx}, which is obtained by the SDSS-II and SNLS collaborations. SN Ia data give measurement of the luminosity distance $D_{\rm L}(z)$ through the measurement of the distance modulus of each SN. The apparent magnitude of SN is
\begin{equation}
m_{\rm{mod}}=5\log_{10} [H_0 D_{\rm{L}}(z)]-\alpha (s-1)+\beta \mathcal{C}+\mathcal{M},
\end{equation}
where the luminosity distance $D_{\rm L}(z)$ is linked to a cosmological model through 
\begin{equation}
D_{\rm L}(z)=\frac{1+z}{H_0}\int_{0}^{z}\frac{dz'}{E(z')},
\end{equation}
$s$ is the stretch measure of the SN light curve shape, $\mathcal{C}$ is the color measure for the SN, $\mathcal{M}$ represents some combination of the absolute magnitude of fiducial SN and the Hubble constant $H_0$. $\alpha$ is stretch-luminosity parameter and $\beta$ is color-luminosity parameter. In this paper, we treat $\alpha$ and $\beta$ as constants. For the studies of time-varying $\beta$ of SN Ia, see e.g. \cite{Wang:2013yja,Wang:2013tic,Wang:2013zca,Wang:2014oga,Zhang:2014ija}.

For a set of $N$ SNe with correlated errors, the $\chi^2$ function is
\begin{equation}
\chi^2_{\rm{SN}}=\Delta \textbf{m}^T\cdot {\textbf{C}}_{\rm SN}^{-1}\cdot \Delta \textbf{m},
\end{equation}
where $\Delta {\bf m} \equiv {\bf m}_B-{\bf m}_{\rm mod}$  is a vector with $N$ components, $m_B$ is the rest-frame peak $B$ band magnitude of SN, and $\textbf{C}_{\rm SN}$ is the $N\times N$ covariance matrix of SN. Here, $N$ denotes the number of SN data points, and for the case of JLA sample, $N=740$.

\subsection{Cosmic microwave background}

For the CMB data, we use the ``Planck distance priors'' derived from the Planck 2015 released data \cite{Ade:2015rim}. The ``distance priors'' include the ``shift parameter" $R$, the ``acoustic scale" $\ell_{\rm A}$, and the ``baryon density" $\omega_{\rm b}$, respectively, defined as
\begin{equation}
R=\sqrt{\Omega_{\rm{m0}}H^2_0}(1+z_\ast)D_{\rm A}(z_\ast),
\end{equation}
\begin{equation}
\ell_{\rm A}=(1+z_\ast)\pi D_{\rm A}(z_\ast)/r_{\rm s}(z_\ast),
\end{equation}
\begin{equation}
\omega_{\rm b}=\Omega_{\rm b0}h^2,
\end{equation}
where $\Omega_{\rm{m0}}$ and $\Omega_{\rm{b0}}$ are the present-day fractional energy densities of dark matter and baryon, respectively. $D_{\rm A}(z_\ast)$ is the proper angular diameter distance at the redshift of the decoupling epoch of photons $z_\ast$. $r_{\rm s}(z_\ast)$ is the comoving size of the sound horizon at $z_\ast$. $D_{\rm A}(z)$ and $r_{\rm s}(z)$ are given by
\begin{equation}
D_{\rm A}(z)=\frac{1}{H_0(1+z)}\int_{0}^{z}\frac{dz'}{E(z')},
\end{equation}
\begin{equation}
r_{\rm s}(z)=\frac{1} {\sqrt{3}} \int_0^{1/(1+z)} \frac{{\rm d}a}
{a^2H(a)\sqrt{1+(3\Omega_{\rm{b0}}/4\Omega_{\gamma0})a}},
\end{equation}
where $\Omega_{\rm{\gamma0}}$ is the present-day fractional energy density of photon. Thus, we have $3\Omega_{b0}/4\Omega_{\gamma0}=31500\Omega_{\rm b0}h^2(T_{\rm{cmb}}/2.7K)^{-4}$, with $T_{\rm cmb}=2.7255K$. $z_\ast$ is given by the fitting formula \cite{Hu:1995en},

\begin{equation}
z_\ast=1048[1+0.00124(\Omega_{\rm{b0}}h^2)^{-0.738}][1+g_1(\Omega_{\rm{m0}}h^2)^{g_2}],
\end{equation}
where
\begin{equation}
g_1=\frac{0.0783(\Omega_{\rm{b0}}h^2)^{-0.238}}{1+39.5(\Omega_{\rm{b0}}h^2)^{0.763}},$$   $$g_2=\frac{0.560}{1+21.1(\Omega_{\rm{b0}}h^2)^{1.81}}.
\end{equation}
The $\chi^2$ function of the CMB data is
\begin{equation}
\chi^2_{\rm CMB}=(\xi^{\rm{obs}}_i-\xi^{\rm{th}}_i)(C^{-1}_{\rm CMB})_{ij}(\xi^{\rm{obs}}_j-\xi^{\rm{th}}_j),
\end{equation}
where $\xi_i=(R,\ell_{\rm A},\omega_{\rm b})$ and $C^{-1}_{\rm CMB}$ is the inverse covariance matrix obtained from the Planck TT+lowP data \cite{Ade:2015rim},
$$C^{-1}_{\rm CMB}=\left(\begin{array}{ccc}
  1 & 0.54 & -0.63 \\
  0.54 & 1 & -0.43 \\
  -0.63 & -0.43 & 1
\end{array}\right).$$

\subsection{Baryon acoustic oscillations}

For the BAO data, we use the measurements of the six-degree-field galaxy survey (6dFGS) at $z_{\rm eff}=0.106$ \cite{Beutler:2011hx}, the SDSS main galaxy sample (MGS) at $z_{\rm eff}=0.15$ \cite{Ross:2014qpa}, the baryon oscillation spectroscopic survey (BOSS) ``LOWZ" at $z_{\rm eff}=0.32$ \cite{Anderson:2013zyy}, and the BOSS CMASS at $z_{\rm eff}=0.57$ \cite{Anderson:2013zyy}.

The spherical average gives us the following effective distance measure $D_{\rm V}(z)$,
\begin{equation}
D_{\rm V}(z)=\left[(1+z)^2D^2_{\rm A}(z)\frac{z}{H(z)}\right]^{1/3},
\end{equation}
where $D_{\rm A}(z)$ is the proper angular diameter distance. The BAO data points we use in this paper are given by the observables like $\xi(z)=r_{\rm s}(z_{\rm d})/D_{\rm V}(z)$, where $z_{\rm d}$ denotes the redshift of drag epoch, whose fitting formula is given by \cite{Eisenstein:1997ik}

\begin{equation}
z_{\rm d}=\frac{1219(\Omega_{\rm{m0}}h^2)^{0.251}}{1+0.659(\Omega_{\rm{m0}}h^2)^{0.828}}[1+b_1(\Omega_{\rm{b0}}h^2)^{b_2}],
\end{equation}
where
\begin{equation}
b_1=0.313(\Omega_{\rm{m0}}h^2)^{-0.419}[1+0.607(\Omega_{\rm{m0}}h^2)^{0.674}],
\end{equation}
\begin{equation}
b_2=0.238(\Omega_{\rm{m0}}h^2)^{0.223}.
\end{equation}
The $\chi^2$ function for BAO is given by
\begin{equation}
\chi^2_{\rm BAO}=\sum\limits_{i=1}^4 \frac{(\xi^{\rm obs}_i-\xi^{\rm th}_i)^2}{\sigma_i^2}.
\end{equation}

\subsection{The Hubble constant}

The precise measurements of $H_0$ will be helpful to break the degeneracy between dark energy parameters \cite{Freedman:2010xv}. For the Hubble constant direct measurement, we use the value given by Efstathiou \cite{Efstathiou:2013via}, $H_0=70.6\pm3.3$ km s$^{-1}$ Mpc$^{-1}$, which is a re-analysis of the Cepheid data of Riess et al \cite{Riess:2011yx}. The $\chi^2$ function for the Hubble constant is
\begin{equation}
\chi^2_{H_0}=\left(\frac{h-0.706}{0.033}\right)^2.
\end{equation}

\section{Results and discussion}\label{sec4}

In this section, we discuss the fitting results of the five interacting holographic dark energy models. We show the constraint results of these models obtained by using the SN+CMB+BAO+$H_0$ data and then make a comparison for them. For comparison, the fitting results of the $\Lambda$CDM model by using the same combination of data are also shown.

\begin{table*}[!htp]\small
\setlength\tabcolsep{8pt}
\caption{Summary of the information criteria results.}
\label{table1}
\centering
\renewcommand{\arraystretch}{1.5}
\begin{tabular}{lccc}
\\
\hline\hline
Model  & $\chi^2_{\rm min}$ & $\Delta$AIC & $\Delta$BIC \\
  \hline

$\Lambda$CDM       & $699.3776$
                   & $0$
                   & $0$
                   \\

HDE                & $704.6058$
                   & $7.2282$
                   & $11.8456$
                   \\

IHDE1              & $700.5253$
                   & $5.1477$
                   & $14.3825$
                   \\

IHDE2              & $702.8318$
                   & $7.4542$
                   & $16.6890$
                   \\

IHDE3              & $702.5593$
                   & $7.1817$
                   & $16.4156$
                   \\

IHDE4              & $700.9179$
                   & $5.5403$
                   & $14.7751$
                   \\

IHDE5              & $700.5221$
                   & $5.1445$
                   & $14.3793$
                   \\
\hline\hline
\end{tabular}
\end{table*}

\begin{table*}[!htp]\small
\setlength\tabcolsep{5pt}
\caption{Fitting results of the models. Best-fit values with $\pm1\sigma$ errors are presented. }
\label{table2}
\renewcommand{\arraystretch}{2}\centering
\begin{tabular}{lcccccc}
\\
\hline\hline
Parameter  & HDE & IHDE1 & IHDE2 & IHDE3 & IHDE4 & IHDE5\\
  \hline
$\Omega_{\rm{m0}}$ & $0.3242^{+0.0081}_{-0.0079}$
                   & $0.3213^{+0.0090}_{-0.0075}$
                   & $0.3225^{+0.0090}_{-0.0070}$
                   & $0.3235^{+0.0077}_{-0.0081}$
                   & $0.3233^{+0.0081}_{-0.0084}$
                   & $0.3224^{+0.0085}_{-0.0073}$
                   \\

$\Omega_{\rm{b0}}$ & $0.0522^{+0.0011}_{-0.0012}$
                   & $0.0518^{+0.0027}_{-0.0023}$
                   & $0.0546^{+0.0021}_{-0.0022}$
                   & $0.0547^{+0.0017}_{-0.0022}$
                   & $0.0526\pm0.0013$
                   & $0.0521^{+0.0013}_{-0.0011}$
                   \\

$c$                & $0.7331^{+0.0354}_{-0.0421}$
                   & $0.8294^{+0.0909}_{-0.0625}$
                   & $0.7538^{+0.0455}_{-0.0430}$
                   & $0.7675^{+0.0444}_{-0.0509}$
                   & $0.7868^{+0.0619}_{-0.0500}$
                   & $0.7983^{+0.0633}_{-0.0543}$
                   \\

$\beta$            &...
                   & $0.0782^{+0.0377}_{-0.0347}$
                   & $0.0092^{+0.0058}_{-0.0070}$
                   & $0.0088^{+0.0048}_{-0.0065}$
                   & $0.0346^{+0.0179}_{-0.0173}$
                   & $0.0958^{+0.0424}_{-0.0464}$
                   \\

$h$                & $0.6565^{+0.0076}_{-0.0068}$
                   & $0.6558^{+0.0070}_{-0.0082}$
                   & $0.6399^{+0.0141}_{-0.0126}$
                   & $0.6394^{+0.0142}_{-0.0102}$
                   & $0.6512\pm0.0077$
                   & $0.6545^{+0.0067}_{-0.0079}$
                   \\
\hline\hline
\end{tabular}
\end{table*}

\begin{figure*}[!htp]
\includegraphics[scale=0.3]{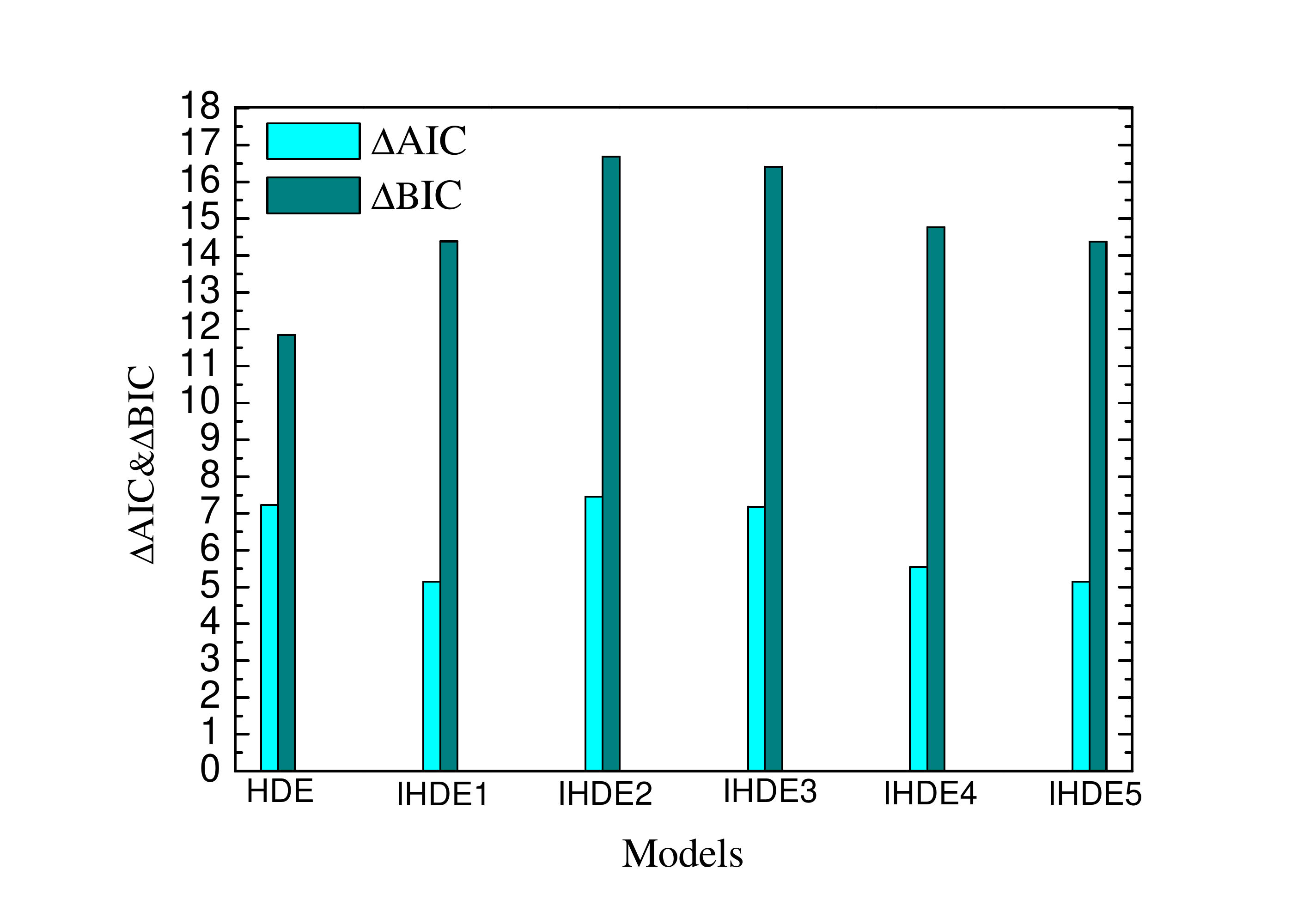}
\centering
 \caption{\label{fig1}Graphical representation of the results of $\Delta$AIC and $\Delta$BIC for the HDE model and the IHDE models.}
\end{figure*}

\begin{figure*}[!htp]
\includegraphics[width=15cm]{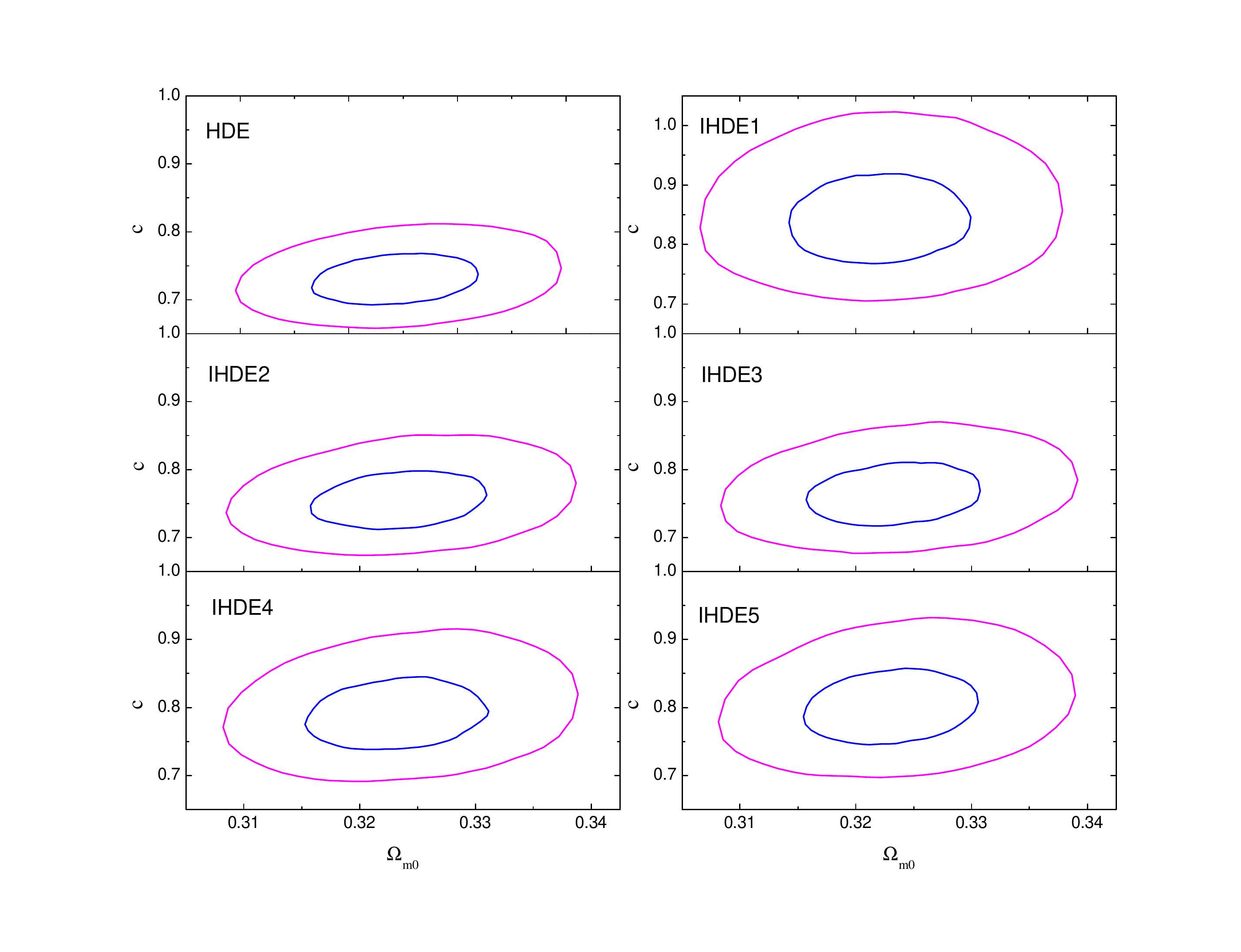}
\caption{\label{fig2}The SN+CMB+BAO+$H_0$ constraints on the HDE model and the IHDE models. The 68.3\% and 95.4\% confidence level contours are shown in the $\Omega_{\rm{m0}}$--$c$ plane. }
\end{figure*}

\begin{figure*}[!htp]
\includegraphics[width=15cm]{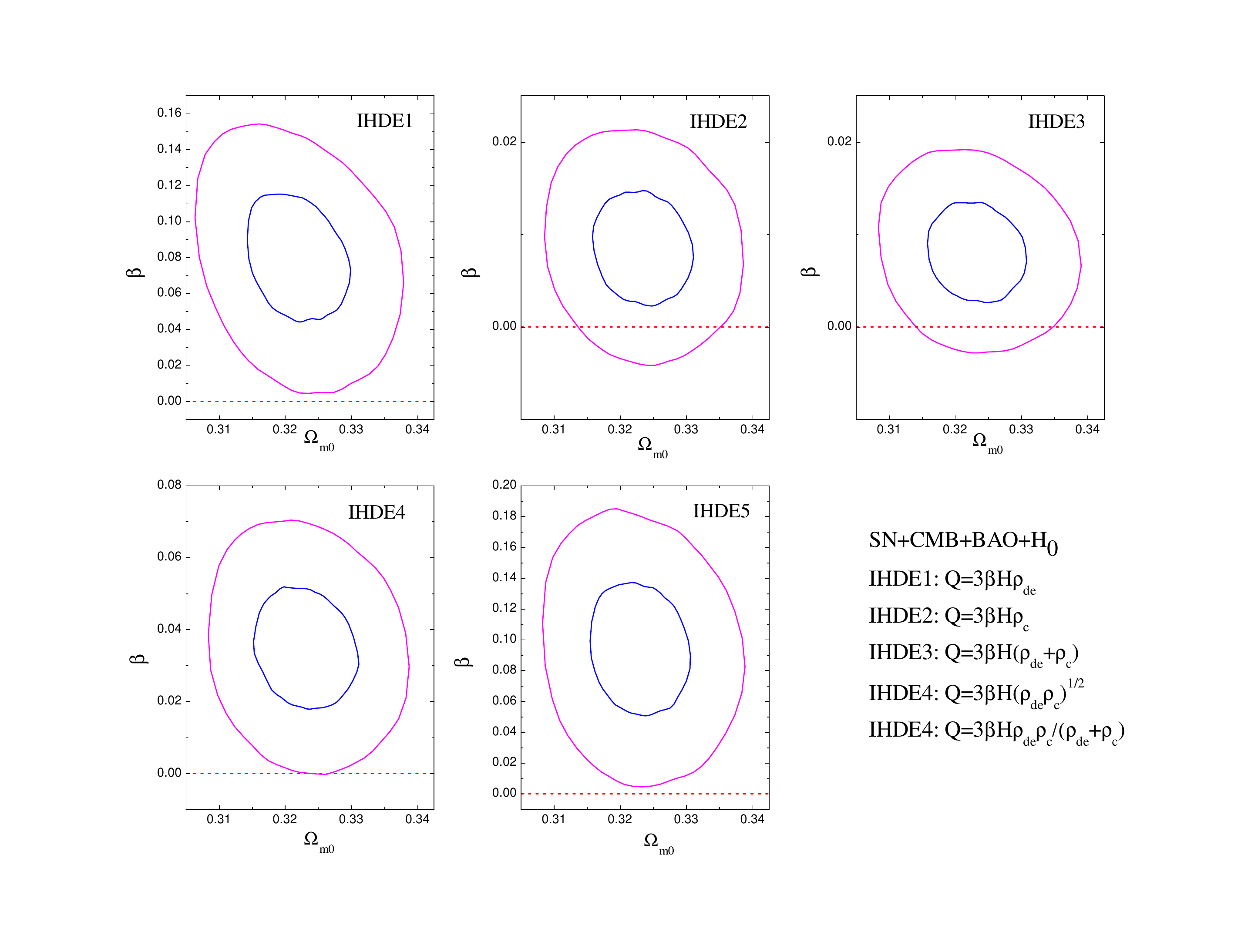}
\caption{\label{fig3}The SN+CMB+BAO+$H_0$ constraints on the HDE model and the IHDE models. The 68.3\% and 95.4\% confidence level contours are shown in the $\Omega_{\rm{m0}}$--$\beta$ plane. The red dashed line denotes the case of $\beta=0$. }
\end{figure*}

\begin{figure*}[!htp]
\includegraphics[width=15cm]{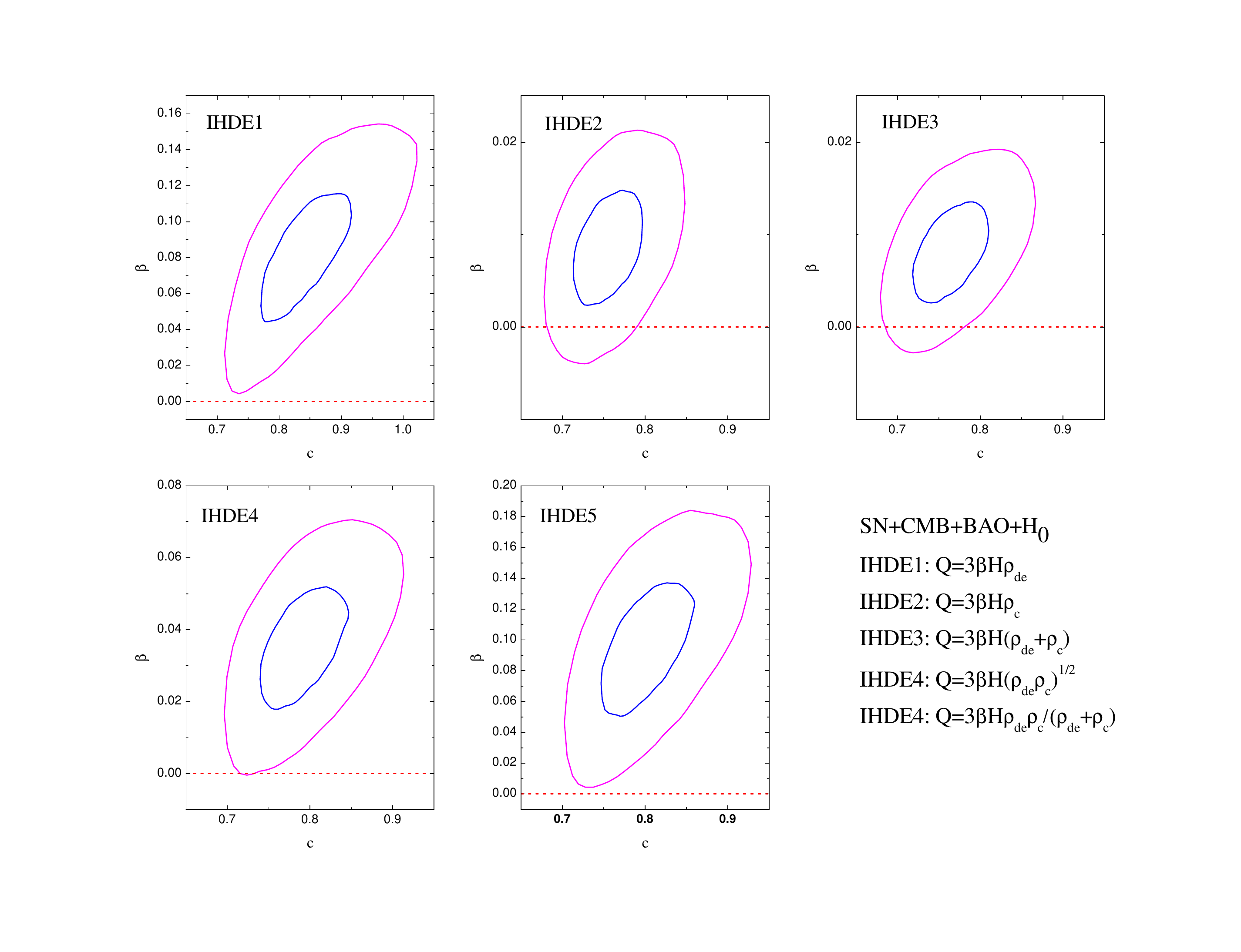}
\caption{\label{fig4}The SN+CMB+BAO+$H_0$ constraints on the IHDE models. The 68.3\% and 95.4\% confidence level contours are shown in the $c$--$\beta$ plane. The red dashed line denotes the case of $\beta=0$. }
\end{figure*}

In table \ref{table1}, $\chi_{\rm min}^2$ and the information criteria values are summarized. The $\Delta {\rm AIC}$ and $\Delta {\rm BIC}$ values are measured with respect to the $\Lambda$CDM model. Since the $\Lambda$CDM model has the lowest AIC and BIC, all the values of $\Delta {\rm AIC}$ and $\Delta {\rm BIC}$ of other models are positive. Among these models, the $\Lambda$CDM model has the least parameters, i.e., less than the HDE model by one parameter and less than the IHDE models by two parameters. We find that although the HDE model has one more parameter than $\Lambda$CDM, it yields a larger $\chi^2$ than the $\Lambda$CDM model. This indicates that in the face of the current accurate observational data, the $\Lambda$CDM model has exhibited remarkable advantage compared with the HDE model in fitting data (see also \cite{Xu:2016grp}). We also find that, even though the IHDE2 and IHDE3 models have two more parameters than $\Lambda$CDM, they still yield larger values of $\chi^2$ than the $\Lambda$CDM model. Of course, these two models have the highest AIC and BIC values among the IHDE models. This indicates that for the interacting holographic dark energy model, the cases of IHDE2 and IHDE3 are not favored by the observational data. Among the interacting models, the IHDE1, IHDE4, and IHDE5 models are more favored by data, with the values of $\Delta {\rm AIC}$ around 5 and $\Delta {\rm BIC}$ around 14. In particular, the IHDE5 model is the best one, with $\Delta {\rm AIC}=5.1445$ and $\Delta{\rm BIC}=14.3793$. The next best one is the IHDE1 model, with $\Delta {\rm AIC}=5.1477$ and $\Delta{\rm BIC}=14.3825$. From this analysis, we find that the $\Lambda$CDM model is much better than the HDE model and the IHDE models in the sense of fitting data. But the meaning of the holographic dark energy model is in that it can provide an interesting mechanism to overcome the theoretical challenges confronted by $\Lambda$CDM. In this work, we focus on the IHDE models and we wish to investigate whether the observational data favor the existence of interaction between dark energy and dark matter in these models.

A graphical representation of the AIC and BIC results is given in figure \ref{fig1}, which directly shows the scores (in the AIC and BIC tests) the models gain.

Before we show the results of parameter estimation, we first discuss the cosmological consequence in the IHDE models. We are interested in the impacts of $c$ and $\beta$ on the EOS of dark energy and the fate of the universe. Note that, no matter if there exists interaction, the EOS of the holographic dark energy always reads
\begin{equation}\label{4.1}
w=-\frac{1}{3}-\frac{2}{3c}\sqrt{\Omega_{\rm de}}.
\end{equation}
In the far future $(z\rightarrow -1)$, it is clear that $\Omega_{\rm de}\rightarrow 1$ and so we still have $w|_{z\rightarrow -1}=-\frac{1}{3}-\frac{2}{3c}$. Hence, we hold the conclusion derived in the HDE model that $c<1$ leads to a big rip future singularity, while for $c>1$ this singularity is avoided.

Though the coupling parameter $\beta$ does not apparanetly enter the expression of $w$ (\ref{4.1}), it can impact the determination of the value of $c$, and thus can affect the evolution of $w$ subtly. In table \ref{table2}, we show the fitting results of the HDE model and the IHDE models. We find that, for all the models, $c<1$ is favored by the observations. In particular, for the HDE model, $c<1$ is favored at the more than 7.5$\sigma$ level by the current data. If the interaction between dark energy and dark matter is considered, the significance of $c<1$ will be decreased. We find that for all the IHDE models, compared to the HDE model, the central value of $c$ is increased and the error range of $c$ is amplified. Among these models, the IHDE1 and IHDE5 model change the estimate of $c$ more evidently. For example, for the IHDE1 model, the statistical significance of $c<1$ becomes about 1.9$\sigma$. In figure \ref{fig2}, we show the 1$\sigma$ and 2$\sigma$ contours in the $\Omega_{\rm{m0}}$--$c$ plane for the HDE model and the IHDE models. In this sense, the interaction between dark energy and dark matter in the holographic dark energy model is helpful in decreasing the significance of appearance of big rip in the future.

\begin{figure*}[!htp]
\includegraphics[width=8cm]{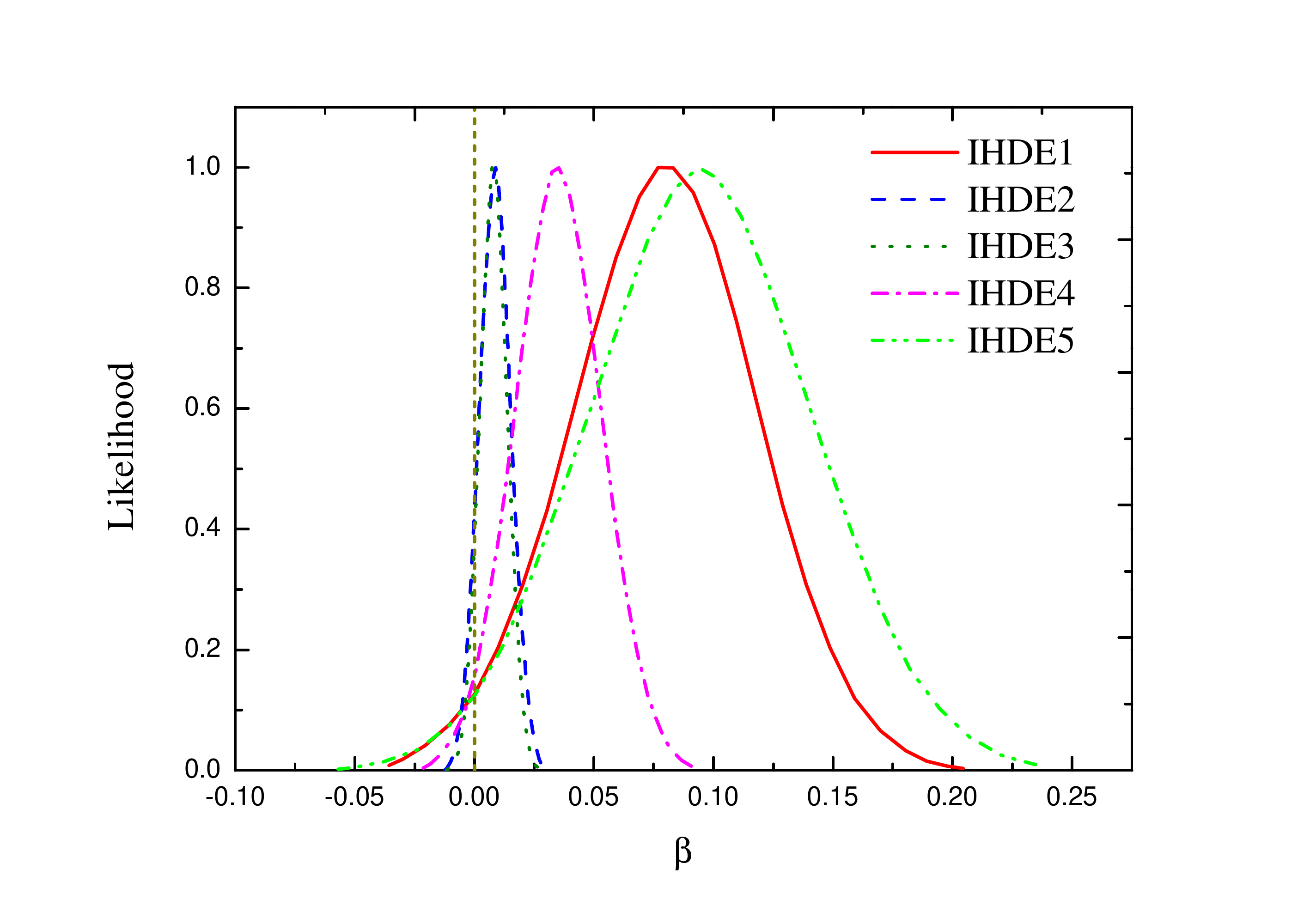}
\includegraphics[width=8cm]{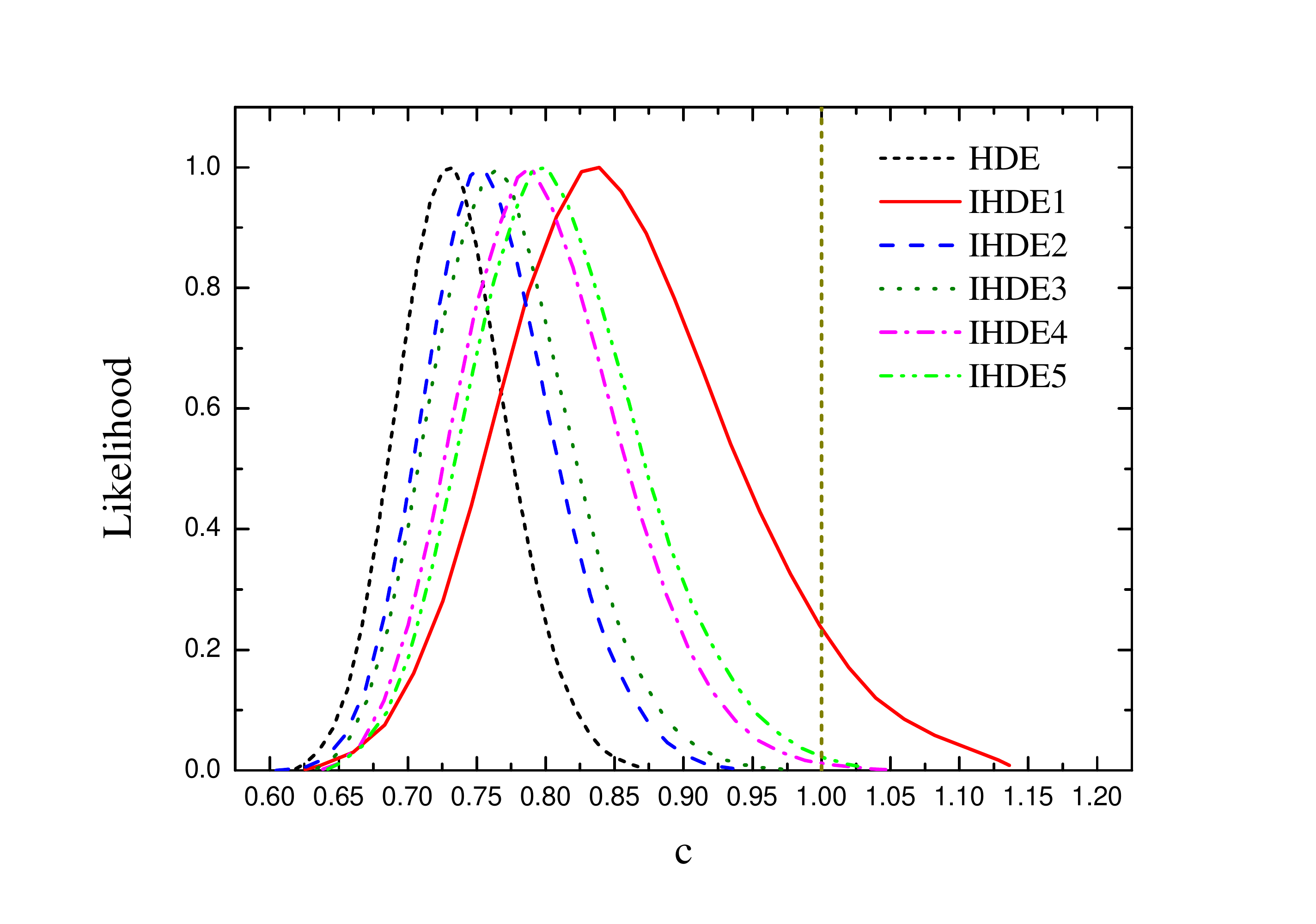}
\caption{\label{fig5} One-dimensional marginalized posterior distributions of parameters $\beta$ (left panel) and $c$ (right panel) for the HDE model and the IHDE models, from the SN+CMB+BAO+$H_0$ data. The dark yellow dashed lines denote the cases of $\beta=0$ (left panel) and $c=1$ (right panel).}
\end{figure*}

In figures \ref{fig3} and \ref{fig4}, we show the 1$\sigma$ and 2$\sigma$ contours in the $\Omega_{\rm{m0}}$--$\beta$ and $c$--$\beta$ planes for the five IHDE models. We find that, for all the models, $\beta$ is in a weak anti-correlation with $\Omega_{\rm{m0}}$, and $\beta$ is in a strong positive correlation with $c$. It is of great interest to find that the detection of $\beta>0$ is at about the 2$\sigma$ level in the IHDE1, IHDE4, and IHDE5 models. In particular, for the IHDE1 model, we have $\beta=0.0782^{+0.0377}_{-0.0347}$, indicating $\beta>0$ at the 2.3$\sigma$ level; for the IHDE5 model, we have $\beta=0.0958^{+0.0424}_{-0.0464}$, indicating $\beta>0$ at the 2.1$\sigma$ level. Since a positive $\beta$ leads to dark energy decaying to dark matter, the risk of holographic dark energy becoming a phantom is decreased, which means that $c$ tends to be increased. This explains why $c$ is positively correlated with $\beta$. In the IHDE1 and IHDE5 models, the detection of $\beta>0$ is at more than 2$\sigma$ level, and thus for these cases $c$ becomes larger, as discussed in the above. In figure \ref{fig5}, we show the one-dimensional marginalized posterior distributions of $\beta$ and $c$ for the models.

\begin{figure*}[!htp]
\includegraphics[width=8cm]{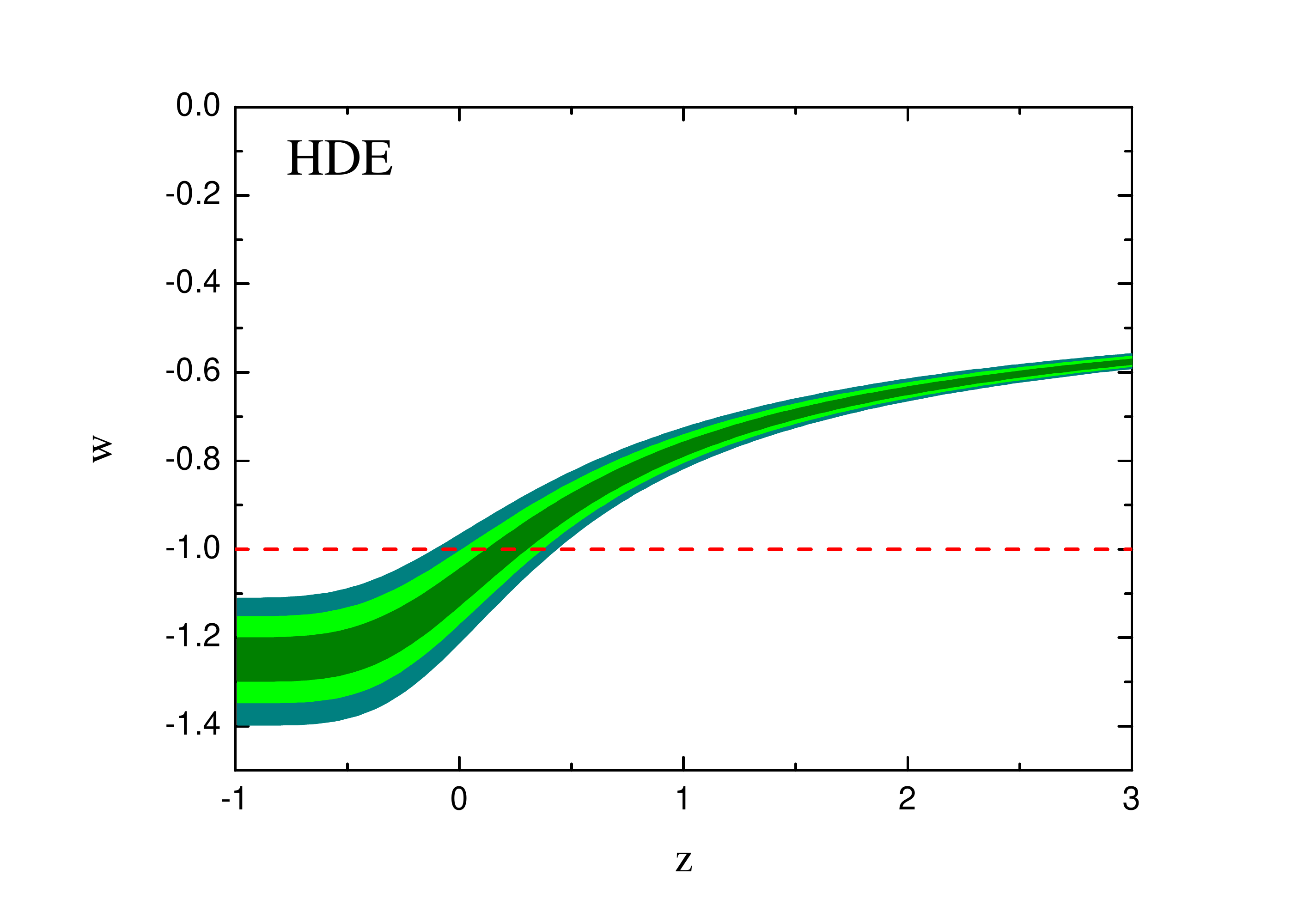}
\includegraphics[width=8cm]{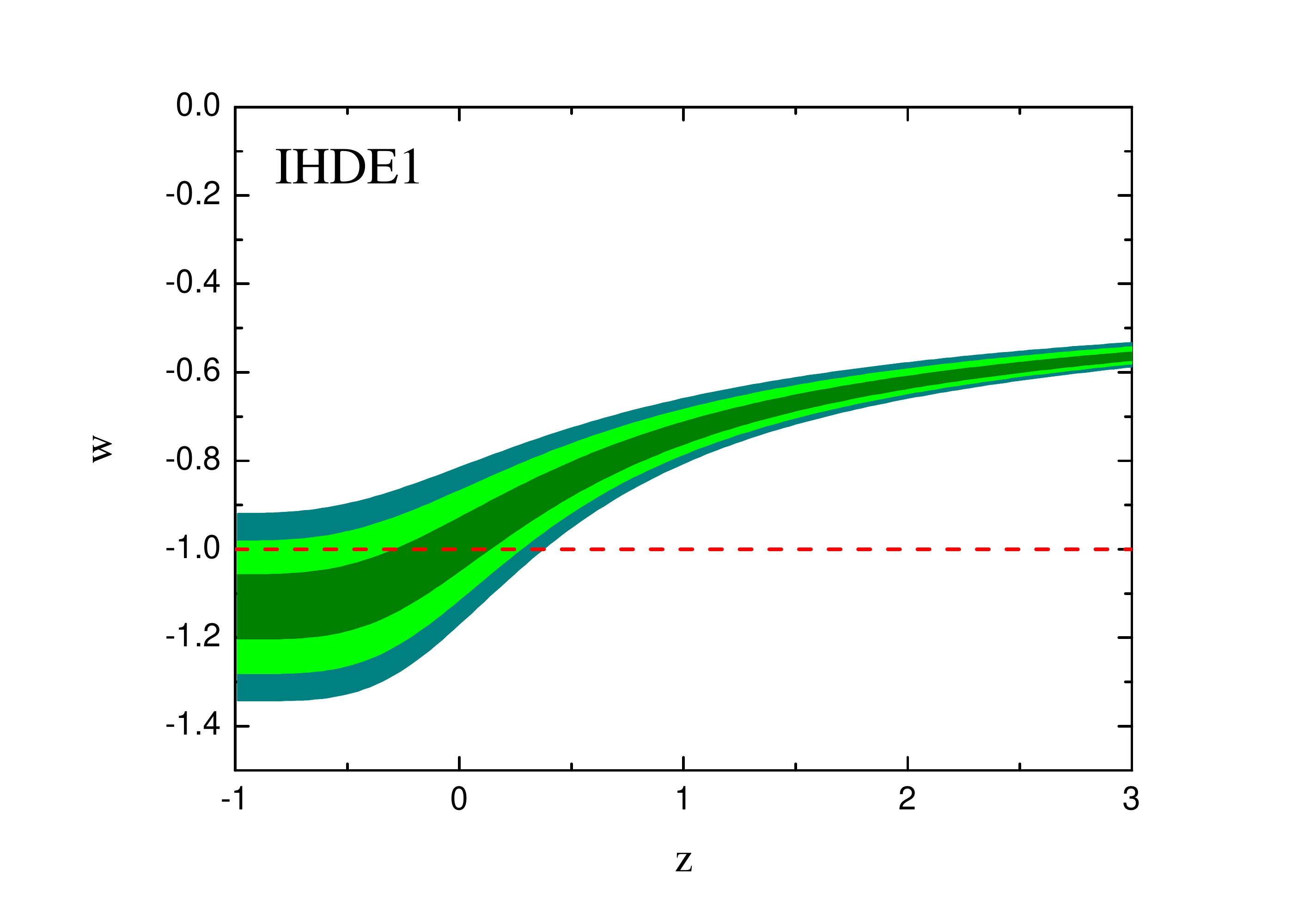}\\
\includegraphics[width=8cm]{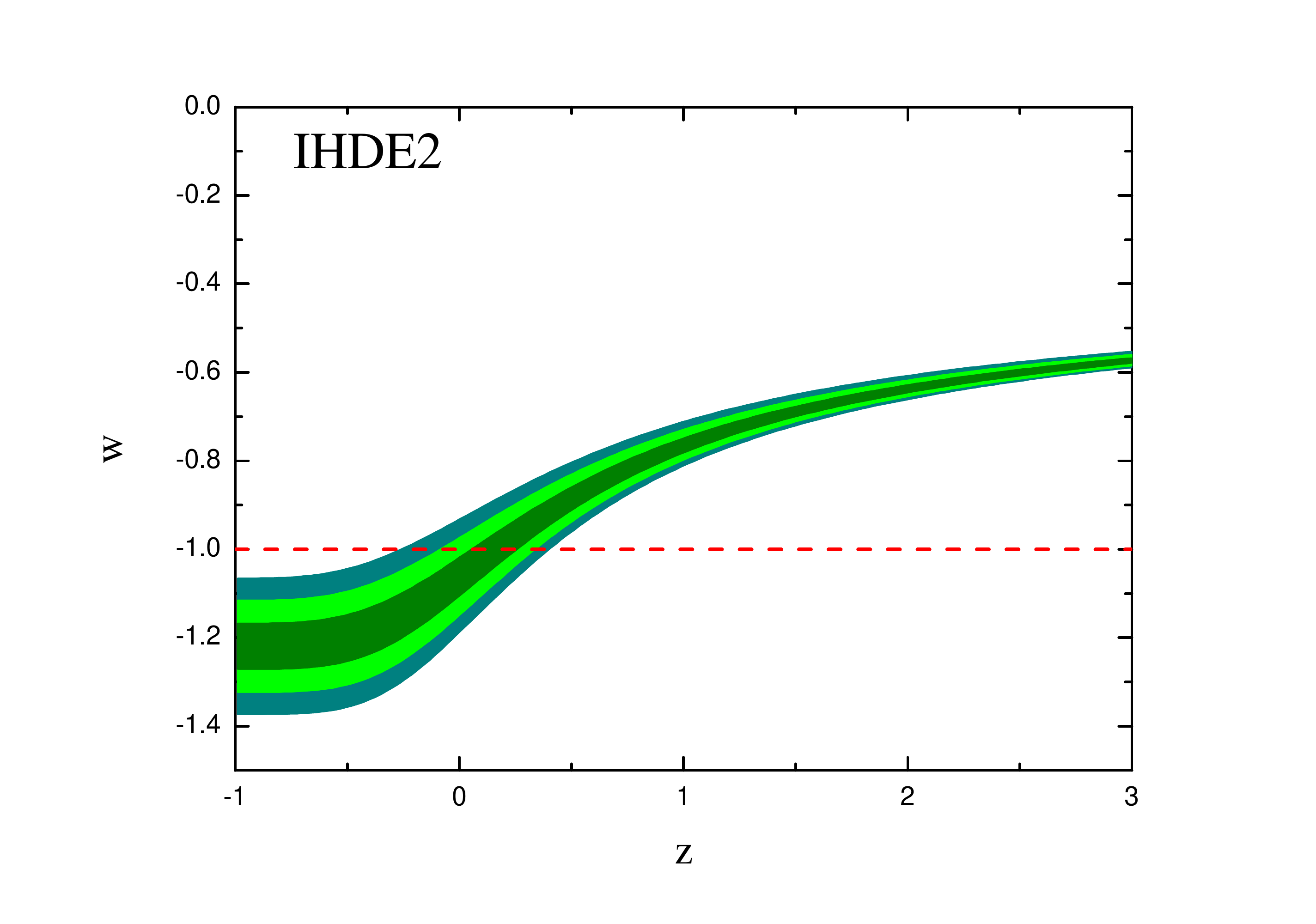}
\includegraphics[width=8cm]{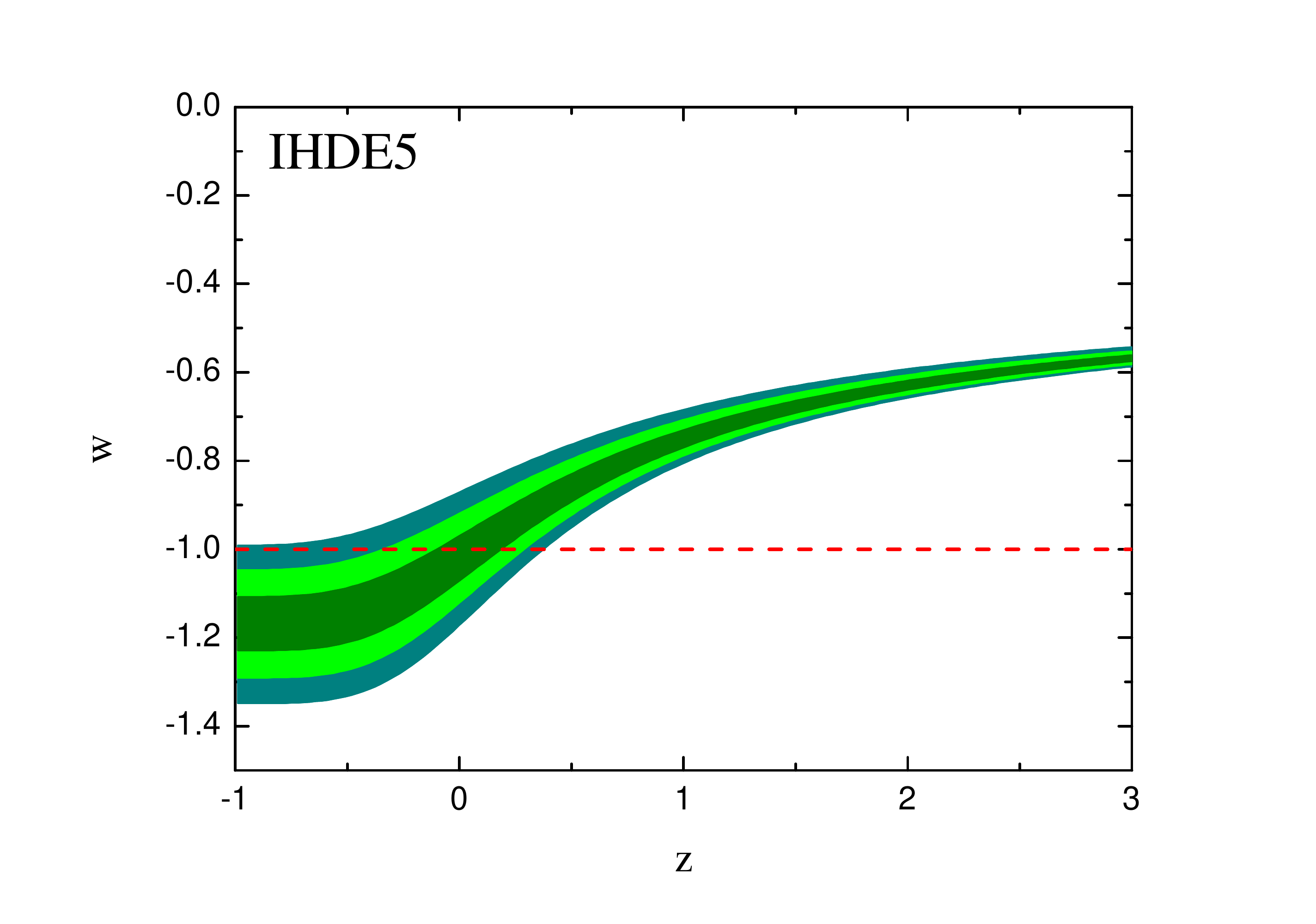}
\caption{\label{fig6} The reconstructed evolution of $w$ (with 1--3$\sigma$ errors) for the HDE, IHDE1, IHDE2, and IHDE5 models. The red dashed line denotes the cosmological constant boundary $w=-1$.}
\end{figure*}

In figure \ref{fig6}, we show the reconstructed evolution of $w$ (with 1--3$\sigma$ errors) for the HDE, IHDE1, IHDE2, and IHDE5 models. We can see that, for the HDE model, the significance of dark energy becoming a phantom is rather high (more than 7$\sigma$ level), and when the interaction is considered, the significance can be decreased. The cases of IHDE1, IHDE2, and IHDE5 are shown as typical examples. For the IHDE2 model, $\beta>0$ is only at the 1$\sigma$ level, and thus the alleviation of a phantom future is limited; but for the IHDE1 and IHDE5 models, $\beta>0$ is at the 2$\sigma$ level, and thus the alleviation is more evident, as the figure shows.

Finally, we note that in this study we have not considered the cosmological perturbations in these models. Since we do not know the physical nature of dark energy, we actually do not know how to properly describe the perturbations of dark energy. Under such circumstances, the usual scheme is to follow the treatment of other components for describing the perturbations of dark energy, i.e., treating dark energy as some fluid and considering it in the framework of hydromechanics under general relativity. In this treatment, we still do not know how the sound waves propagate in the dark energy fluid, and thus we need to impose a rest-frame sound speed for dark energy by hand, which sometimes leads to divergence of dark energy perturbations. For example, it is well-known that the perturbation divergence will happen at the point of $w$ crossing $-1$ \cite{Zhao:2005vj}. It was also found in \cite{Valiviita:2008iv} that in the models of interacting dark energy, for some regions in the parameter space, a kind of early-time super-horizon perturbation divergence also appears. To avoid such instabilities, a parametrized post-Friedmann (PPF) framework for interacting dark energy was established \cite{Li:2014eha,Li:2014cee} (which is an updated version of the original PPF \cite{Fang:2008sn}). Using the PPF approach, the perturbations of dark energy can be considered appropriately, and the observations of structure growth (such as weak lensing and redshift space distortions) can also be considered in the cosmological fits. But in this study, for economical reason, we do not consider the cosmological perturbations in our calculations, and we only use the observations of distance information to constrain the models. A recent work \cite{Herrera:2016uci} shows that a reconstruction method can be used to avoid the undesirable instabilities in the interacting dark energy models, but this treatment will lead to a modification for the corresponding background model, and the parameter estimation would also be changed accordingly. Thus, here we wish to remind the reader that the constraint results obtained in this paper should be treated with caution. Actually, a further step is to investigate the IHDE models within the PPF framework by considering both observational data of expansion history and structure growth.

\section{Summary}\label{sec5}

We have studied the direct, non-gravitational interaction between dark energy and dark matter in the holographic dark energy model. We considered five typical IHDE models: the IHDE1 model with $Q=3\beta H\rho_{\rm{de}}$, the IHDE2 model with $Q=3\beta H\rho_{\rm c}$, the IHDE3 model with $Q=3\beta H(\rho_{\rm{de}}+\rho_{\rm c})$, the IHDE4 model with $Q=3\beta H\sqrt{\rho_{\rm{ de}}\rho_{\rm c}}$, and the IHDE5 model with $Q=3\beta H\frac{\rho_{\rm{de}}\rho_{\rm c}}{\rho_{\rm{ de}}+\rho_{\rm c}}$. We investigated the current status of observational constraints on these models after the 2015 data release of the Planck mission. The observational data we used in this paper include the JLA compilation of SN Ia data, the Planck CMB distance priors data, the BAO data, and the $H_0$ direct measurement.

We have made a comparison for these five IHDE models by employing the information criteria and we found that, for fitting the current data, the IHDE5 model is the best one, the IHDE1 model is the next best one, and the IHDE2 model is the worst one. That is to say, in the framework of holographic dark energy, the $Q=3\beta H\frac{\rho_{\rm{de}}\rho_{\rm c}}{\rho_{\rm{ de}}+\rho_{\rm c}}$ model is most favored by current data, and so this model deserves deeper investigation in the future; the $Q=3\beta H\rho_{\rm{de}}$ model is also a good model; and the $Q=3\beta H\rho_{\rm c}$ model is relatively not favored by the current data.

We found that, within the framework of holographic dark energy, the interaction between dark energy and dark matter can be detected at more than 2$\sigma$ significance. For example, for the IHDE1 model, we have $\beta=0.0782^{+0.0377}_{-0.0347}$, indicating $\beta>0$ at the 2.3$\sigma$ level; for the IHDE5 model, we have $\beta=0.0958^{+0.0424}_{-0.0464}$, indicating $\beta>0$ at the 2.1$\sigma$ level. Since a positive $\beta$ leads to dark energy decaying to dark matter, the result of $\beta>0$ will affect the parameter estimate of $c$, i.e., it tends to make $c$ become larger. We found that $c$ is indeed positively correlated with $\beta$ in the parameter estimates from observations. We have discussed the related issues of evolution of dark energy and fate of the universe.

\acknowledgments

We acknowledge the use of {\tt CosmoMC}. We thank Yun-He Li, Yue-Yao Xu, and Ming-Ming Zhao for helpful discussions.
This work is supported by the Top-Notch Young Talents Program of China, the National Natural Science Foundation of China (Grant No.~11522540), and the Fundamental Research Funds for the Central Universities (Grant No. N140505002).



\end{document}